\documentclass[useAMS,usenatbib]{mn2e}
\usepackage{times}

\title[Observations of Arp 220]{Continuum and spectral line observations of the OH Megamaser galaxy Arp 220}
\author[Rovilos et al.]{E. Rovilos$^{1}$\thanks{E-mail: erovilos@jb.man.ac.uk}, P. J. Diamond$^{1}$, C. J. Lonsdale$^{2}$, C. J. Lonsdale$^{3}$ and H. E. Smith$^{4}$\\
$^{1}$Jodrell Bank Observatory, University of Manchester, Macclesfield, Cheshire, SK11 9DL, U.K.\\
$^{2}$MIT Haystack Observatory, Off Route 40, Westford, MA 01886, U.S.A.\\
$^{3}$Infrared Processing and Analysis Center, California Institute of Technology, 100 22, Pasadena, CA 91125, U.S.A.\\
$^{4}$Center for Astrophysics and Space Sciences, University of California, San Diego, La Jolla, CA 92093-0424, U.S.A.}

\begin{document}

\pagerange{\pageref{firstpage}--\pageref{lastpage}} \pubyear{2002}

\maketitle

\label{firstpage}

\begin{abstract}
We present MERLIN observations of the continuum (both 1.6 and
5 GHz) and OH maser emission towards Arp 220. The correct spatial
configuration of the various components of the galaxy is revealed. In
the eastern component the masers are shown to be generally coincident
with the larger-scale continuum emission; in the west, the masers and
continuum do not generally arise from the same location. A velocity
gradient ((0.32$\pm$0.03)\,km\,s$^{-1}$pc$^{-1}$) is found in the
eastern nuclear region on MERLIN scales; this gradient is three times
smaller than that seen in HI and implies that the OH gas lies inside
the HI. A re-analysis of previously presented global VLBI data
\citep{b9} reveals a very high velocity gradient
((18.67$\pm$0.12)\,km\,s$^{-1}$pc$^{-1}$) in one component, possibly
the site of a heavily obscured AGN.
\end{abstract}

\begin{keywords}
masers -- galaxies:individual:Arp220 -- galaxies:active -- radio continuum:galaxies -- radio lines:galaxies.
\end{keywords}

\section{Introduction}

Ultra-luminous infrared galaxies (ULIRGs) are so named because their
infrared luminosity is greater than the luminosity in all the other
wavelengths combined ($L_{IR}>10^{12}L_{\odot}$). The source of such
copious emission is thought to be the enormous amount of molecular
dust in their central regions, which obscures visible and other
wavelengths and produces infrared light through heating. The dust
grains are excellent absorbers of short wavelength radiation emitted
from newly formed OB stars, so a high infrared luminosity is a tracer
of star formation. In many cases however active galactic nuclei and
Seyfert-like nuclei can play an important role. A review of the
properties of this kind of galaxy can be found in \citet{b14}.

One of the most interesting properties of ULIRGs is that they tend to
be merging systems, as it takes a strong merger of dust rich galaxies
to generate the observed high density star formation. The proportion
of the mergers among ULIRGs gets larger when the infrared luminosity
gets higher, and is 100\% in luminosities above
$10^{12}L_{\odot}$. (See \citet{b14} and references therein). Many
of them also have very powerful OH maser emission (e.g. \citet{b21})
in which the luminosity exceeds that of known galactic masers by a
factor of at least 10$^{6}$, hence the term ``megamaser''.

Arp~220 is one of the most luminous infrared galaxies with
$L_{IR}=1.4\times10^{12}L_{\odot}$ \citep{b19} and provides a superb
example of the starburst phenomenon, demonstrated graphically by the
discovery of several radio supernovae \citep{b18}. Starburst phenomena
are used by \citet{b17} and \citet{b8} to explain the X-ray properties
of Arp~220, although the existence of a heavily obscured AGN is not
ruled out. Recent CHANDRA results \citep{b5} show that the hard X-Ray
emission in Arp~220 is confined in a sub-kiloparsec scale region, in
contrast to other starburst galaxies and its spectrum shows that the
hard X-rays are more likely to be produced by one or more low
luminosity, heavily obscured AGN, X-Ray binaries or Ultra Luminous
X-Ray sources rather than the supernovae. Therefore the co-existence
of such structures is still plausible.

The morphology of Arp~220 is also an open issue. At optical
\citep{b16} as well as at radio wavelengths (eg. \citealt{b11}) it
exhibits a double structure with tidal tails and dust lanes which
leads to the assumption that it is the result of a recent merging
phenomenon. Studies of CO, HI and the OH maser emission enable the
determination of its velocity structures. On balance, the current data
appear to favor a model of two counterrotating disks as parts of a
larger scale disk \citep*{b13,b10}, but a single warped circumnuclear
disk model can also be fit to the data \citep{b7}. In the case of the
related galaxy IIIZw35 the existence of evidence for such a disk,
where the OH maser is located, was recently revealed \citep{b12} so a
search for a similar structure in Arp~220 would be useful.

The very powerful maser emission is perhaps the most striking property
of Arp~220. It was first discovered by \citet{b4} while searching for
OH absorption in galaxies with known HI absorption. The masers were
thought to arise in regions extending to several hundreds of parsecs
\citep{b1}. However, higher resolution studies \citep{b6,b9} revealed
a dual component structure: there is a diffuse component which
consists of approximately one third of the 1667 MHz OH emission plus
almost all of the 1665 MHz OH emission with a physical extent of some
hundred parsecs as suggested by the Baan model; but there is also a
compact component consisting only of the 1667 MHz line with a physical
extent of just a few parsecs. These compact masers show high velocity
widths, even in this very small physical extent \citep{b9}. This means
that either they come from a region with a volume of a few pc$^{3}$ or
they represent long, thin filaments with their orientation along the
line of sight. In the first case there has to be a hidden infrared
source in that region to provide the essential pumping photons, or the
optical depth of this region has to be large enough to trap the
infrared photons, if the pumping is indeed radiative and not
collisional due to shock fronts from the nearby supernovae. If
filamental structure was present to any great extent we would have to
explain why the filaments' orientation is only along the line of
sight, so that we cannot see elongated structures \citep{b20}.
Probably we are only seing filaments which lie on the line of sight
due to the gain path needed for a maser to be detected.

In this paper we present new MERLIN OH spectral line and continuum
data, and compare it with previously presented global VLBI data
\citep{b18,b9}.

\section{Observations and Data Reduction}

In this section we describe the MERLIN and global VLBI observations
of both line and continuum emission

\subsection{MERLIN Observations}
\subsubsection{1.6 GHz Spectral Line Data}

Data were taken on 23 April 1998 with seven MERLIN antennas including
the 76\,m Lovell telescope observing with left-hand circular
polarization only. The dataset included observations of the target
source Arp~220 (spending 9 hours and 20 minutes on source) and the
phase reference 1511+238, as well as two scans on the point source
calibrator 2134+004 and one scan on the flux calibrator 3C286. The
observations covered a bandwidth of 4 MHz and were correlated to
produce 256 channels, each with a bandwidth of 15.625 KHz.

The initial editing and gain elevation corrections, as well as
preliminary amplitude calibration were made using MERLIN specific
software. Then, the data were transferred into NRAO's Astronomical
Image Processing System (AIPS) for bandpass calibration and phase
corrections. Phase referencing was performed using the data on
1511+238. The data were then self-calibrated starting from a point
source model and the solutions were applied to the Arp~220 data. After
that line free channels were averaged and an image of this continuum
was produced in `difmap'. There may be modest contamination of the
continuum image by weak 1665 MHz OH emission but tests have shown this
to be minimal. We attempted to image the 1665 MHz emission but it was
too diffuse even on MERLIN scales; it was detected only on short
baselines, shorter than $0.35\,M\lambda$. Several iterations of
self-calibration were performed on the continuum data and the
solutions obtained were then applied to the multi-frequency dataset
and the data averaged for 60 seconds. The final spectral line dataset
was generated by subtracting the continuum emission using UVLIN and
then averaging every two channels together, resulting in a velocity
resolution of 5.6\,km\,s$^{-1}$.

\subsubsection{1.6 GHz Continuum Data}

Continuum data were taken on 23 June 1998 with six MERLIN antennas
using right hand circular polarization. The dataset includes
observations of the target source Arp~220 (spending 4 hours on source)
and the phase reference 1511+238, as well as two scans on the point
source calibrator 2134+004 and two scans on the flux calibrator 3C286.
16 channels were generated by the correlator, each with a bandwidth of
1 MHz, the central channel (8) being set at a frequency of
1666.00 MHz.

Local MERLIN software was used to perform the gain elevation
corrections and preliminary amplitude calibration and
editing. Bandpass calibration was performed and AIPS was used for
further calibration. After flux calibration, the phase reference
source was used for phase calibration, and three rounds of
self-calibration were performed on 1511+238 starting from a
point-source model.  The solutions were copied and applied to all the
data, channels 2 to 15 were averaged and a first image of Arp~220 was
produced. Starting from this image, we performed four iterations of
self calibration on the Arp~220 data before creating the final image.

\subsubsection{5 GHz Continuum Data} 

5 GHz data were taken using 6 MERLIN antennas in full polarization on
29 May 2000. The dataset consisted of observations of Arp~220
(spending 7 hours on source) and the phase reference 1551+239 as well
as five scans on the point source 2134+004 and two scans on the flux
calibrator 3C286.  Again, 16 channels were produced by the correlator
each with a bandwidth of 1 MHz, the central channel being set at a
frequency of 4.995 GHz.

Standard procedures were followed for the data reduction, the phase
corrections from the phase reference source were applied to the
Arp~220 data and after self-calibration the final image was produced
using Stokes I.

\subsection{VLBI Observations}

The VLBI data used for comparison are those discussed in our previous
papers, \citet{b18} and \citet{b9}. The data were taken in November
1994 and discussion of the calibration is contained with the papers.

\section{Results}
\subsection{Continuum}
\subsubsection{MERLIN}

\begin{figure}
\setlength{\unitlength}{1cm}
\begin{picture}(1,7)
\put(-1.5,0){\includegraphics{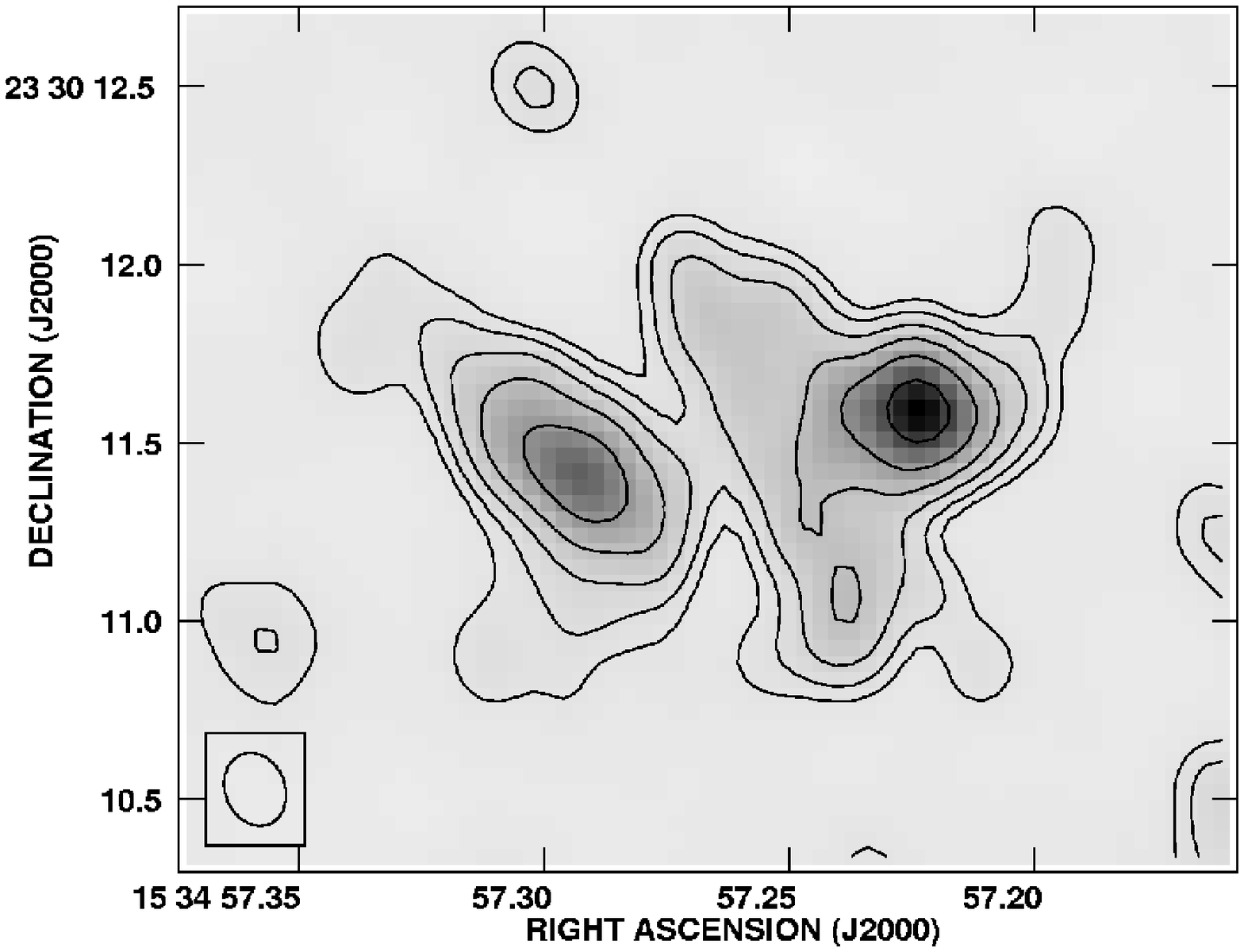}}
\end{picture}
\caption{\small The 1.6 GHz image of Arp~220 from the MERLIN continuum
          data. The peak flux is 44.604\,mJy\,beam$^{-1}$ and the
          contour levels are at 1, 2, 4, 8, 16 and 32\,mJy. The beam
          size is shown at the bottom left corner of the image and is
          209.97$\times$167.48\,mas (77.4$\times$61.7\,pc) and the off
          source noise level is 055\,mJy\,beam$^{-1}$.}
\end{figure}

The 1.6 GHz (June 1998) MERLIN continuum image of Arp~220 is shown in
Figure 1: A double component structure is present, with the western
and the eastern components separated by $\sim$1\,arcsec ($\sim$370\,pc
if we assume a distance of 76\,Mpc;
H$_{0}$=75\,km\,s$^{-1}$\,Mpc$^{-1}$) and at the same positions as the
1.3\,mm \citep{b13} and the 4.83 GHz (\citealt{b3}; this paper)
continuum peaks. The integrated flux densities are (91.9$\pm$0.9)\,mJy
and (111.2$\pm$0.8)\,mJy for the eastern and the western components
respectively. We should note here that the component boundaries are
somewhat arbitrary, so the overall diffuse emission plays a role in
deriving these numbers: the values given are lower limits. The total
integrated flux density of the galaxy at 1.6 GHz is (280$\pm$40)\,mJy,
consistent with that derived at 1.4 GHz by \citet{b10}
(285.4$\pm$14.3\,mJy).

In Figure 1 we can see that the eastern component is elongated in a
northeast - southwest direction and a `spur' of emission is visible on
the northwest of the western component. This is also detected by
\citet{b10} at 1.4 GHz and by \citet{b3} in the 4.8 GHz continuum.  A
component southeast of the western component is also detected as part
of a more extended emission pattern, possibly part of the extended
emission detected by the VLA \citep{b2}. The peak flux of this region
is 8.72\,mJy\,beam$^{-1}$, whereas the off-source noise level is
0.55\,mJy\,beam$^{-1}$.

In Figure 2 we can see the 5 GHz image of Arp~220 where again the
double component structure is present, although more extended
structure is not detected. The peaks of the two components are
separated by (1.066$\pm$0.004)\,arcsec and their integrated fluxes are
(33.6$\pm$1.0)\,mJy and (60.7$\pm$0.8)\,mJy for the east and west
respectively, while the off-source noise level is
0.4\,mJy\,beam$^{-1}$. If we compare these fluxes with these of the
1.6 GHz data we can derive the spectral indices of these two continuum
components to be:\\
$\alpha_{e}=-0.90\pm0.03$ \\
$\alpha_{w}=-0.542\pm0.013$ \\
The errors shown here are statistical. There might be systematic
effects that cause larger errors. The spectral indices derived here
are similar to the spectral index (for the whole nucleus) assumed by
\citet{b11} (-0.6 to -0.9). This is a characteristic of synchrotron
radiation (see next paragraph) though it cannot give us a safe
indication of the exact origin of the emission or its power source.

\begin{figure}
\setlength{\unitlength}{1cm}
\begin{picture}(1,6)
\put(-1.5,0){\includegraphics{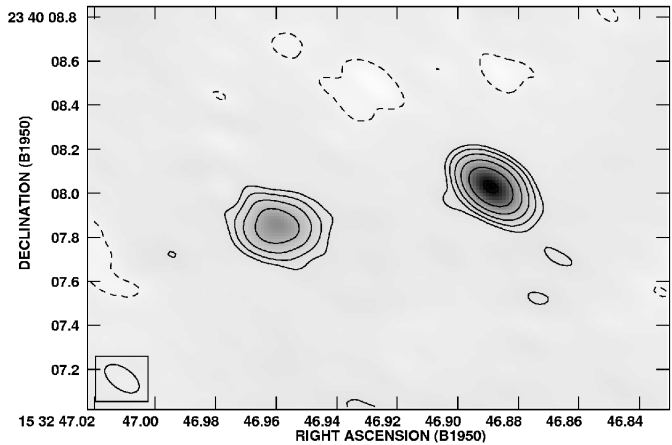}}
\end{picture}
\caption{\small The 5 GHz image of Arp~220 from the MERLIN continuum
          data. The peak flux is 34.6\,mJy\,beam$^{-1}$ and the
          contour levels are at 1, 2, 4, 8, 16, and 32\,mJy, as in
          Figure 1. The beam size is shown at the bottom left corner
          of the image and is 178.42$\times$95.16\,mas
          (65.7$\times$35.1\,pc). The off-source noise level is
          0.4\,mJy\,beam$^{-1}$.}
\end{figure}

\subsubsection{Global VLBI}

The results from the continuum VLBI dataset have been previously
presented by \citet{b18}, where the discovery of probable radio
supernovae (RSNe) is reported. There are about a dozen RSNe (November
1994) in the western nuclear region and possibly one or two in the
eastern, without any other obvious compact continuum component.
(Figures 1a and 1b respectively in \citet{b18}) The sum of the flux
densities of the observed supernovae in the western region is
7.87\,mJy and in the eastern is 0.78\,mJy including the marginally
detected ones. If we compare these with the flux densities observed
with MERLIN it is safe to assume that the MERLIN and VLBI continuum
detections are the results of different processes; the large-scale
MERLIN continuum emission probably comes from synchrotron acceleration
of free electrons (possibly generated in the supernova explosions),
while the VLBI emission probably arises from the interaction between
the explosion material and the stellar wind generated before the star
exploded.

\subsection{Spectral Line}
\subsubsection{MERLIN}

The naturally weighted velocity maps of Arp~220 (after continuum
subtraction) can be seen in Figure 3. There are three components
present, one in the east and two in the west (northwest and
southwest). The eastern component is more redshifted than those in the
west, suggesting an overall velocity gradient along the east-west
axis. This is in agreement with the suggested presence of an overall
rotating disk \citep{b15,b13,b10}. The off-source noise level is
3.2\,mJy\,beam$^{-1}$.

\begin{figure*}
\setlength{\unitlength}{1cm}
\begin{picture}(40,12.1)
\put(1,-0.2){\includegraphics{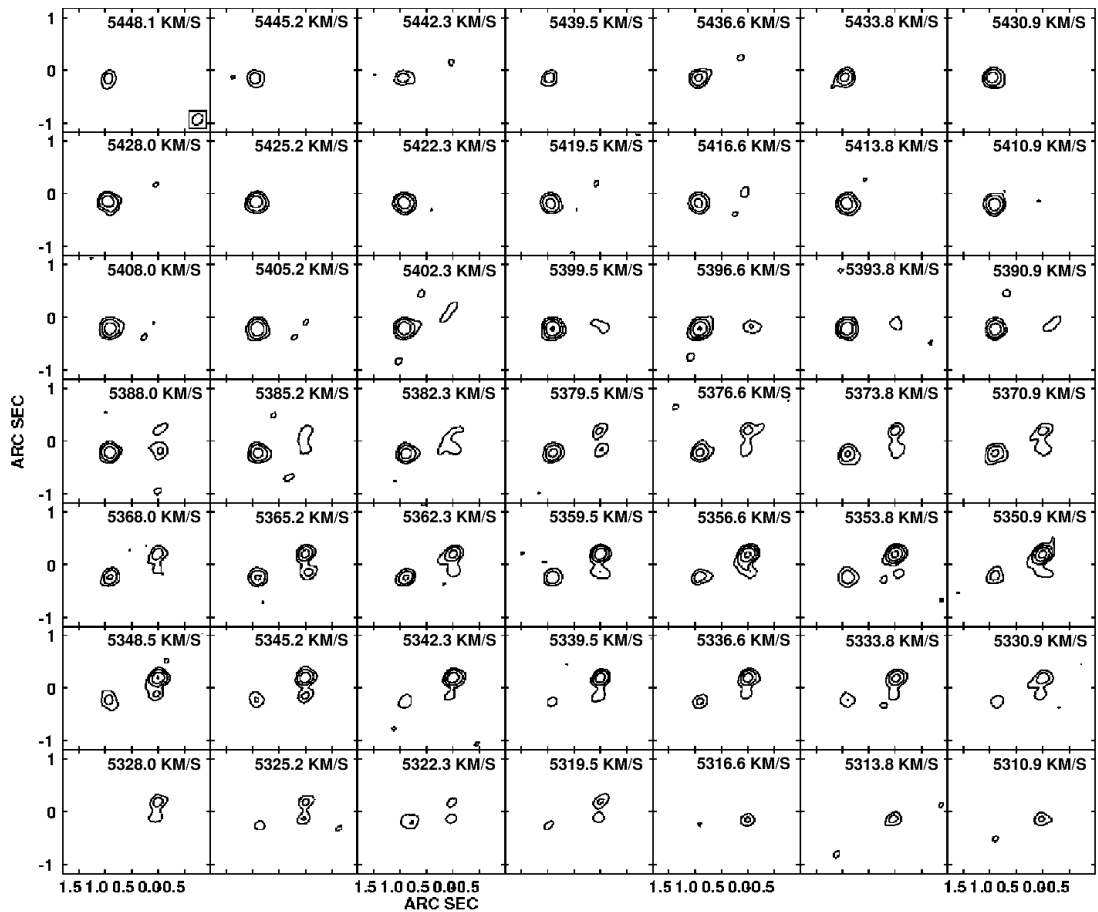}}
\end{picture}
\caption{\small The natural weighted images of Arp 220 from the MERLIN
          spectral line data. The peak integrated flux is
          166.73\,mJy\,beam$^{-1}$. The beam size is shown at the
          bottom right corner of the first image and is
          229.96$\times$182.75\,mas (84.7$\times$67.3\,pc). The noise
          level of each image is 3.2\,mJy\,beam$^{-1}$.}
\end{figure*}

In Figure 4 we can see the spectral line emission (naturally weighted
and averaged over the velocity range of the OH emission) superimposed
on the continuum emission.  This continuum image (in greyscale) is not
the same as that in Figure 1, but was generated from the continuum
portion of the spectral line dataset. The error in relative position
between the continuum and the spectral line image is less than a
milliarcsecond but the noise in this continuum image is significantly
greater than that obtained from the wideband continuum data
(0.83\,mJy\,beam$^{-1}$). The eastern portion of the spectral line
emission is seen to be at the same position as the continuum, but the
peak of the western component of the continuum emission lies between
the two regions of spectral line emission. The northwestern line
component is at the same position as the compact VLBI component and
the southwestern emission corresponds to the previously observed
diffuse component \citep{b9}. The eastern component spectrum is broad
and has a double peak, which implies that regions with different
velocities are superimposed (visible with the VLBI) to produce the
component detected with MERLIN; their spectra are blended together.

\subsubsection{Global VLBI}

\citet{b9} presented the results of the VLBI spectral line dataset,
where they find both a diffuse and a compact component in each of the
eastern and the western regions. In Figure 5 we can see the spectral
line VLBI image of Arp 220 (in contours) with respect to the continuum
emission (in greyscale). Both images are convolved with a taper of
10\,M$\lambda$ to include both regions in one map, so they are of
degraded spatial resolution. In the western region we can detect the
maser in regions where we have continuum emission due to the
supernovae, but the majority of the compact maser emission is detected
in regions north and south of the supernovae, where no compact
continuum is detected. In the case of the eastern region there is no
clear compact continuum emission, only hints of possible radio
supernovae; the maser does not follow the pattern seen in the west.
Close-ups of the two nuclear regions are shown in contours (the same
as in figures 1 and 2 of \citet{b9}) with full resolution. (Beam size
9.8$\times$2.5\,mas). Spectra of different regions are also shown. One
interesting point to note is the fourth spectrum (lower left in Figure
5), which shows absorption. We should note that there are problems
with imaging diffuse emission with the global VLBI due to the lack of
short uv spacings, we are however confident of some absorption being
present. This suggests that the OH radical is present even in regions
with no maser emission. We also detect part of the 1665 MHz emission,
clearly seen in the two diffuse regions (spectra 5 and 8) and in the
eastern compact region (spectra 1 and 2). To determine the origin of
this emission we used tighter boxes to determine the spectra; the
1665 MHz emission then fades dramatically with respect to the 1667 MHz
emission. This suggests that it is diffuse and comes from the overall
area, a result first reported in \citet{b9}.

%

\begin{figure*}
\setlength{\unitlength}{1cm}
\begin{picture}(40,27.5)
\put(-0.2,-0.20){\includegraphics{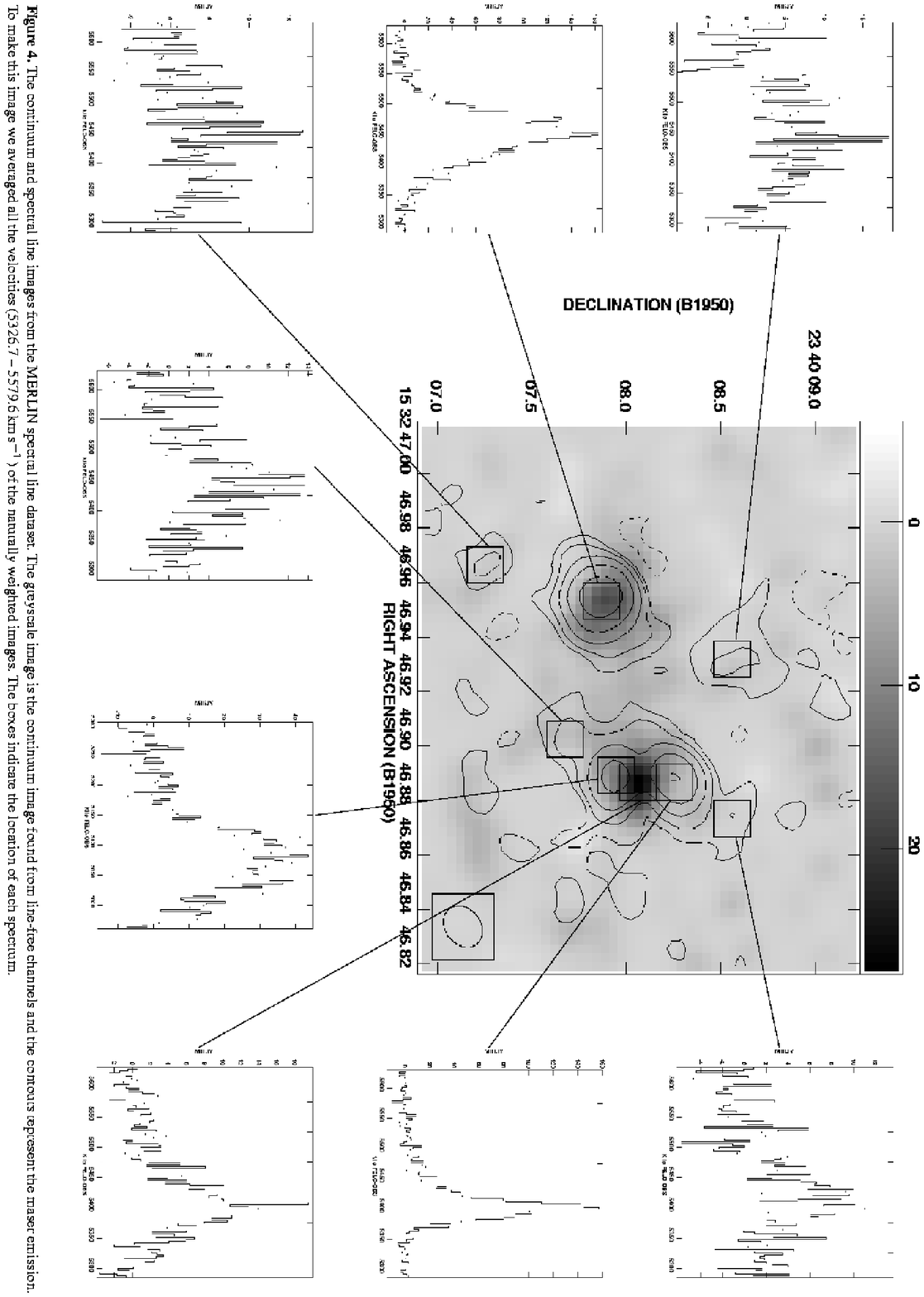}}
\end{picture}
\end{figure*}

\begin{figure*}
\setlength{\unitlength}{1cm}
\begin{picture}(40,27.5)
\put(-1,-0.20){\includegraphics{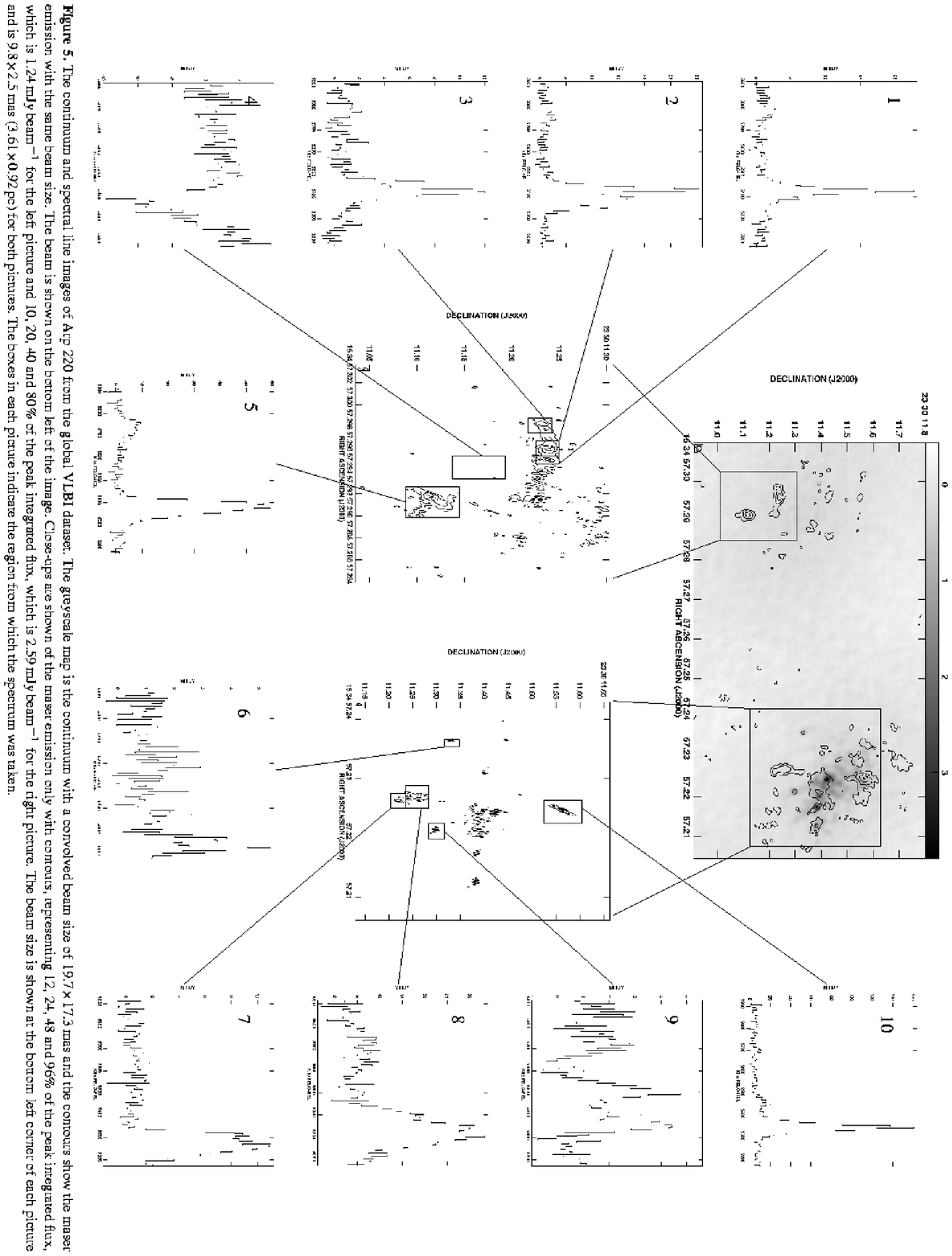}}
\end{picture}
\end{figure*}

\section{Discussion}

\begin {table*}
\begin{tabular}{|c|c|c|c|}                                   \hline\hline
                        &        & $\delta$RA\,(arcsec)   & $\delta$DEC\,(arcsec)  \\ \hline
 NW maser – W cont   & MERLIN & 0.011$\pm$0.006        & 0.191$\pm$0.004        \\ \cline{2-4}
                        & VLBI   & 0.0024$\pm$0.0007      & 0.1394$\pm$0.0007      \\ \hline
 NW maser – SW maser & MERLIN & -(0.012$\pm$0.011)     & 0.311$\pm$0.007        \\ \cline{2-4}
                        & VLBI   & -(0.0335$\pm$0.0007)   & 0.3032$\pm$0.0016      \\ \hline
 W cont – SW maser   & MERLIN & -(0.022$\pm$0.008)     & 0.12$\pm$0.02          \\ \cline{2-4}
                        & VLBI   & -(0.0359$\pm$0.0012)   & 0.1638$\pm$0.0017      \\ \hline
 NW maser – E maser  & MERLIN & -(0.993$\pm$0.006)     & 0.339$\pm$0.004        \\ \cline{2-4}
                        & VLBI   & -(1.0710$\pm$0.0002)   & 0.3252$\pm$0.0006      \\ \hline\hline
\end{tabular}
\caption{\small The positional differences between the individual
          components in both the MERLIN and VLBI spectral line and
          continuum emission. }
\end{table*}

\subsection{Positions of the Masers}

Any discussion of positions must be prefaced by the comment that since
we performed self calibration, information about the absolute
positions of the components is lost. The conclusions we make below are
based on the relative positions of the components. We have extremely
accurate relative positions (sub-mas) since we are able to obtain low
sensitivity images of the continuum emission from the spectral line
datasets; these can be used to tie down the position of the high
sensitivity continuum image with respect to the maser positions.

It is clear that in both MERLIN and VLBI maps the continuum emission
does not coincide with the maser peaks. Table 1 lists the measured
offsets between various maser and continuum components in the MERLIN
and VLBI images.  (We chose the brightest supernova as the location of
the VLBI western continuum peak and the brightest maser component
(area of spectrum 2) as the peak of the spectral line emission in the
east).

From this table we can assume that the northwestern MERLIN component
of the maser emission is at the same position as the compact
northwestern VLBI maser component and the southwestern maser MERLIN
component is at the same position as the diffuse southwestern VLBI maser
component. In this case, the brightest radio supernova is on the
northern part of the western MERLIN continuum emission,
$\sim$0.05\,arcsec offset from its center. In the eastern region, the
VLBI maser emission is situated on the border of the MERLIN maser
emission. Since the MERLIN maser and bulk of the continuum coincide
within $\sim$0.07\,arcsec, and the possible supernovae are situated in
the area between the two VLBI maser components \citep{b18}, they lie
in the area of the MERLIN continuum emission.

Unfortunately, this registration of the different components in
Arp~220 requires a revision of the alignment suggested by \citet{b10}
(see their figure 8). We are confident that our alignment is correct,
as we extracted both the continuum and line images of Figure 4 from
the same dataset performing the same self calibration. This may, to
some degree, affect their conclusions regarding the location of the
supernovae within their tilted disc model.

\subsubsection{Eastern Region}

If we assume that the MERLIN 1.6 GHz continuum emission coincides with
the 1.4 GHz continuum emission \citep{b10}, then all 1.6 GHz
continuum, 1.4 GHz, 1.3\,mm \citep{b13} and 4.8 GHz \citep{b3} come
from the same area, as does the HI absorption \citep{b10} and CO
emission \citep{b13}. Weak, tentatively detected supernovae are also
situated there.

The MERLIN maser emission coincides with the continuum in the eastern
component. The VLBI maser emission lies in regions with no obvious
compact continuum background; it does not cover all the area covered
by the MERLIN maser but is instead situated on the edge of this area
having a face-on disk-like appearance. It is probably saturated; the
lower limit of the amplification factor is 110 for the northern and 65
for the southern region (for the 1667 MHz line). However, the OH
radical is present throughout the whole area, as exemplified by the
detection of OH absorption.

\subsubsection{Western Region}

In the western region, the MERLIN 1.6 GHz, the 1.4 GHz, 4.8 GHz and
1.3\,mm continuum emission come from the same area as the CO
emission and HI absorption. The strong supernovae are situated there,
close to the northern border, probably assisting the young OB stars in
heating the dust grains. The position of the OH maser emission is
however different. Both MERLIN and VLBI demonstrate that it arises
from regions north and south of the continuum. The lower limit for the
amplification factor (derived from the VLBI data) is 125 for the
compact northern region, 45 for the southern diffuse region and 90 for
the compact region on its southwest border, so that the two compact
regions are very likely to be saturated. OH is present at some level
throughout the western region; we can see some maser emission lying in
front of two supernovae northeast and northwest from the diffuse maser
region. However, the amplification factors at the line peaks in front
of the supernovae are very small, 4.6$\pm$0.7 and 3.3$\pm$0.3
respectively. This means that they lie in the area of diffuse,
low-amplifying OH gas, which we only can detect in those regions with
significant continuum radiation.  Observations with better sensitivity
will be able to detect more sources of this kind and so study the
properties of the diffuse OH in some detail.

\setcounter{figure}{5}
\begin{figure}
\setlength{\unitlength}{1cm}
\begin{picture}(4,7.5)
\put(-0.5,0){\includegraphics{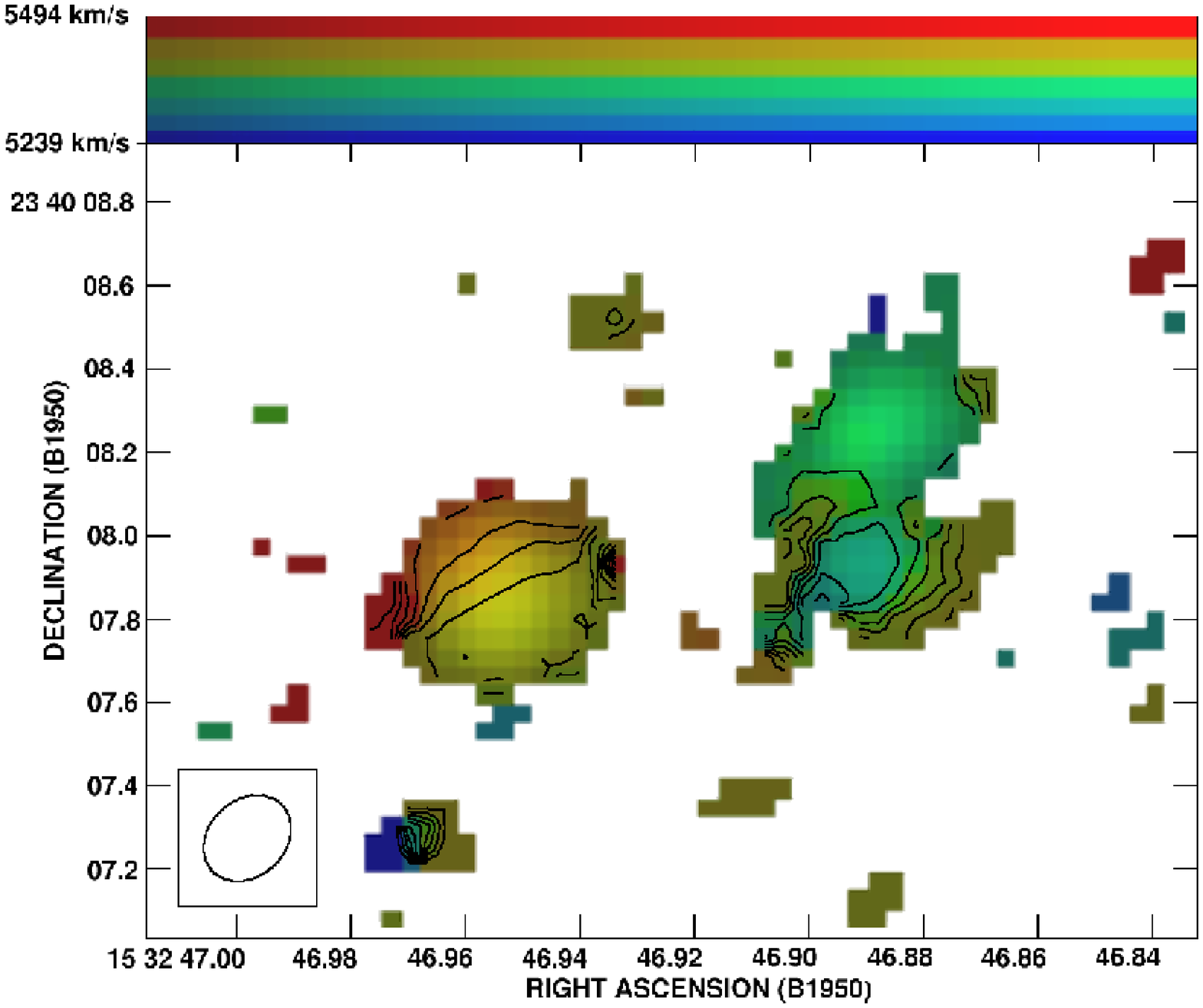}}
\end{picture}
\caption{\small The moment map of Arp~220 from the MERLIN spectral
          line data. The intensity represents the flux density and the
          colors the velocity field, the reddest being the most
          redshifted. The contours represent the velocity field and
          they range from 5305 to 5475\,km\,s$^{-1}$ with spacings of
          10 km\,s$^{-1}$. The area selected to plot the velocity
          field is where the integrated flux is more than
          15\,mJy\,beam$^{-1}$.}
\end{figure}

\begin{figure}
\setlength{\unitlength}{1cm}
\begin{picture}(6,5.5)
\put(-0.6,0){\includegraphics{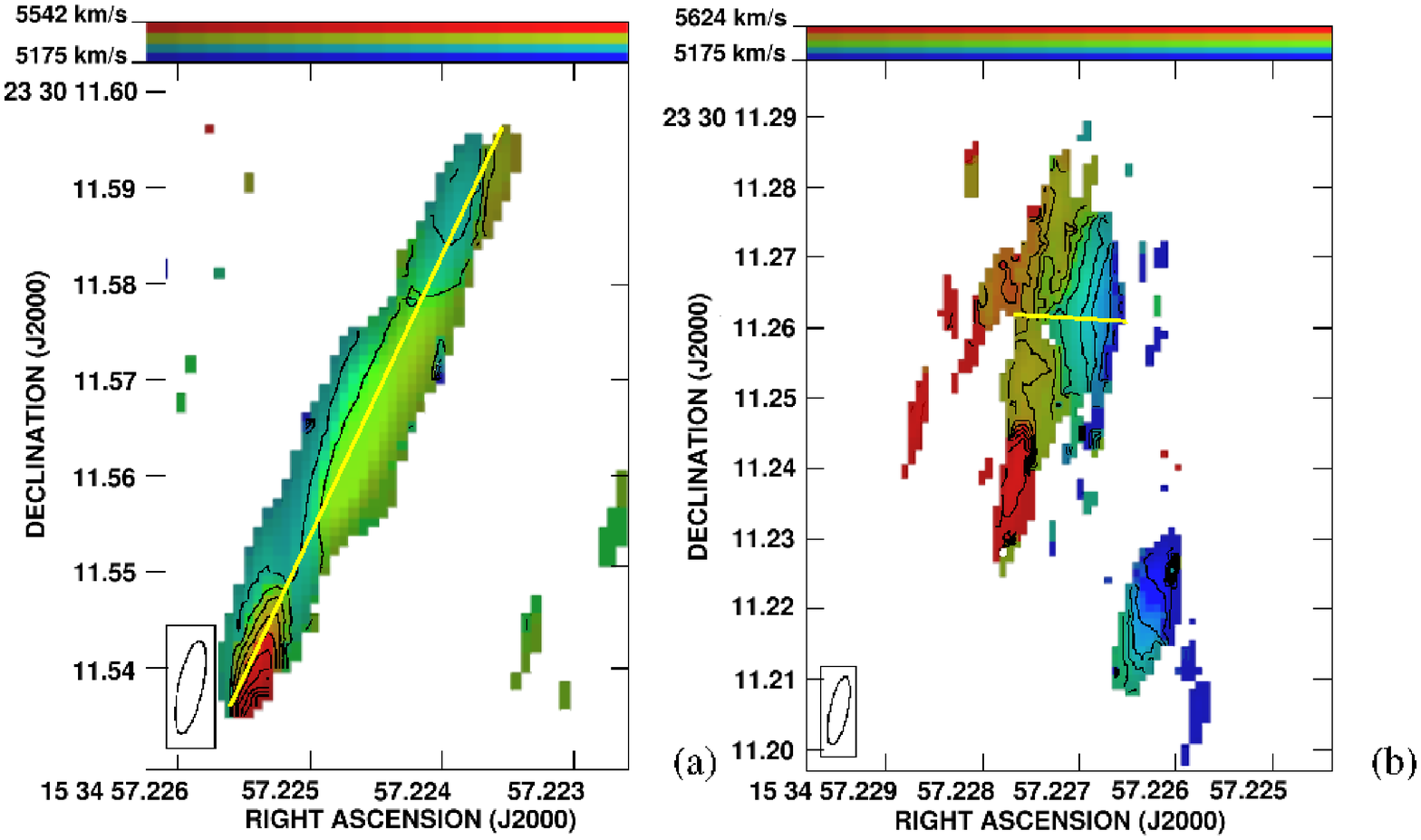}}
\end{picture}
\caption{\small The moment maps of the western region of Arp~220 from
          the VLBI dataset. The left picture (a) represents the
          northern compact region and the right (b) the southern
          diffuse region. The contours for both pictures are
          12.5\,km\,s$^{-1}$ apart, representing a velocity field from
          5250 to 5612.5\,km\,s$^{-1}$. Only data whose
          integrated flux is greater than 1.5\,mJy\,beam$^{-1}$ are selected.}
\end{figure}

\subsection{Velocity Gradiens}

Figure 6 shows the moment map taken from the MERLIN data. There is a
clear velocity gradient of (0.32$\pm$0.03)\,km\,s$^{-1}$pc$^{-1}$ in
the eastern region, at a position angle of 35.5\degr. \citet{b10} find
(1.01$\pm$0.02)\,km\,s$^{-1}$pc$^{-1}$ for the HI at a PA of
$\sim$55\degr. If we assume that the OH arises from a circular
rotating disc, its radius is (79$\pm$5)\,pc and its dynamical mass is
(12$\pm$3)$\times$10$^{6}$M$_{\odot}$. This is smaller by a factor of
90 than the mass of 1.1$\times$10$^{9}$M$_{\odot}$ calculated from HI
absorption. This large difference may be explained if the OH gas lies
in the inner part of the disk and the HI and CO arise in the outer
part. This disc may be warped, explaining the difference in position
angle between the OH and HI gradients.

\citet{b10} and \citet{b13} find a clear velocity gradient in HI and
CO in the western region of Arp~220, in a roughly horizontal
direction. In the MERLIN image of the OH (Figure 6) we can see some
velocity structure in the southwestern component, although a gradient
is not clearly visible and in the northwestern component no gradient
is detectable. However, the VLBI moment maps of Figures 7a and 7b (of
the northern and southern region respectively) show that the southern
component has a roughly E-W velocity gradient of
(18.67$\pm$0.12)\,km\,s$^{-1}$pc$^{-1}$ (Figure 8). The region from
which this gradient arises is on the southern border of the region
that contains the HI and CO gradient and its diameter is
(5.9$\pm$0.5)\,pc. If this gas was in rotation, the enclosed mass
would be (1.7$\pm$0.4)$\times$10$^{7}$M$_{\odot}$. The velocity
gradient we measure is $\sim$20 times greater and the mass is $\sim$10
times smaller than that measured using HI absorption
(0.83$\pm$0.02)\,km\,s$^{-1}$pc$^{-1}$ and
1.7$\times$10$^{8}$M$_{\odot}$ respectively. A plausible explanation
is that the OH is confined in a region of much smaller physical extent
than the HI, in the center of the molecular disk. However, in
comparison to the eastern region the mass is concentrated in the
centre of the disk and that gives rise to the very high velocity
gradient. Such a disk is however not well confined, so we cannot
derive reliable results about the mass or determine how the density
decreases with distance.

\begin{figure}
\setlength{\unitlength}{1cm}
\begin{picture}(6,7.3)
\put(-0.7,0){\includegraphics{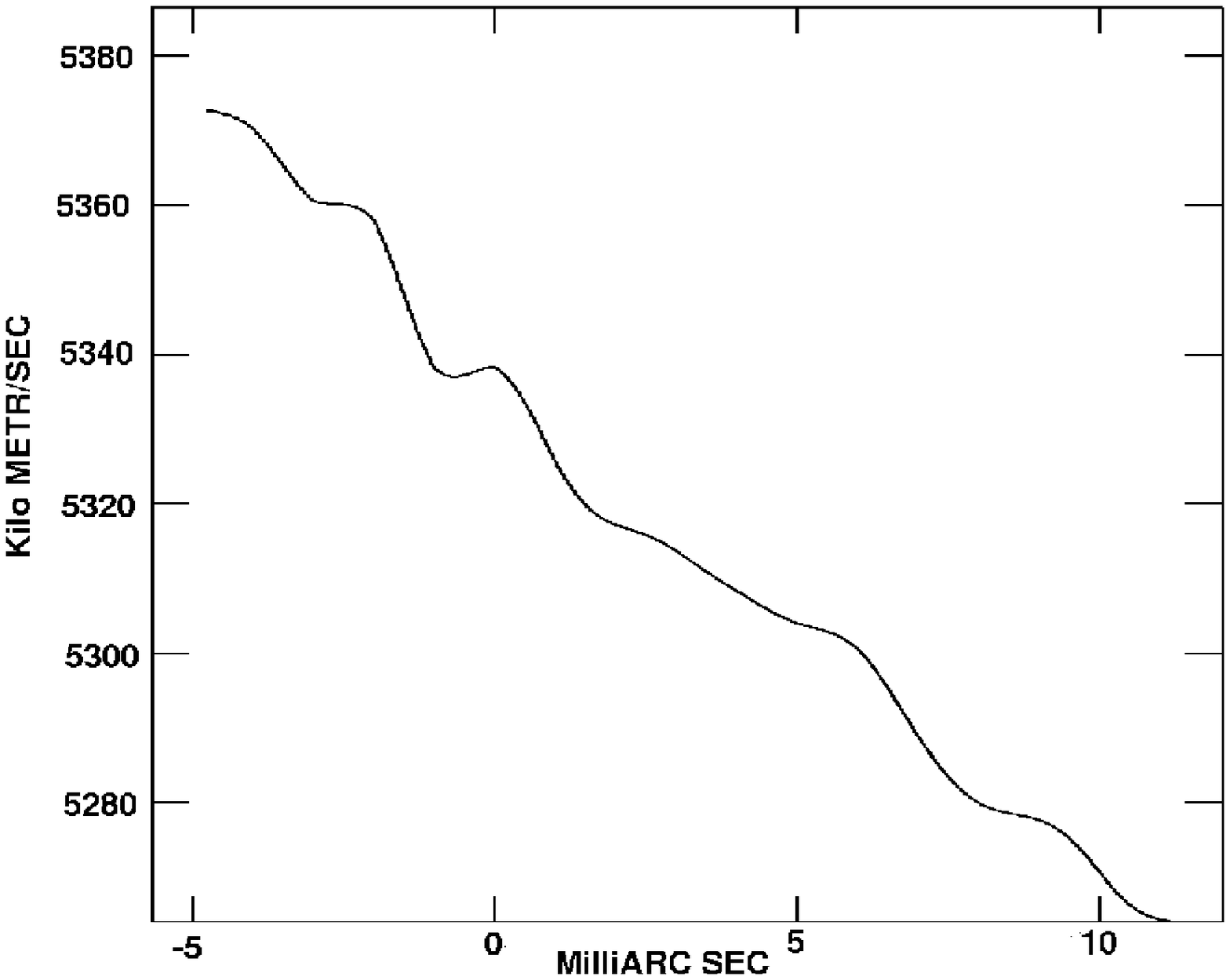}}
\end{picture}
\caption{\small The velocity gradient of the southwestern region of
          Arp~220. The points from which the velocities were taken lie
          on the yellow line of Figure 7b with a position angle of
          85.2\degr and no smoothing has been applied. The velocity
          gradient was derived using the least square method and is
          (18.67$\pm$0.12)\,km\,s$^{-1}$pc$^{-1}$.}
\end{figure}

The northwestern component shows a very strange velocity structure. On
the MERLIN scales no obvious structure is detected, whereas the VLBI
structure can be seen in Figure 9. If we accept that the systemic
velocity of this region is $\sim$5355\,km\,s$^{-1}$ (ie. the velocity
of the central part of Figure 7a), then the region just outside and on
both sides of the centre is blueshifted and it becomes redshifted as
we go further out. Two possible explanations are proposed to explain
this structure, the first being that the masing gas experiences spiral
outflow within a cone, thus exhibiting an expanding helix structure
(Figure 10). The problem with this configuration is that the velocity
structure is symmetrical, so that one helix would be a mirror image of
the other. Alternatively, there might be outgoing and ingoing
spherical shock-waves producing the redshifted and blueshifted
velocities (Figure 11). The observation of an elongated structure
would be due to beaming effects.

To distinguish between these two proposed or other possible structures
and investigate the nature of the central region higher spatial
resolution observations have to be obtained, or a better
signal-to-noise ratio has to be achieved. We will then be able to
examine the velocity structure further away from the central point to
see if it continues to be redshifted or instead becomes blueshifted
again, something which would favor the outflow scenario.

\begin{figure}
\setlength{\unitlength}{1cm}
\begin{picture}(6,8.6)
\put(-0.5,0){\includegraphics{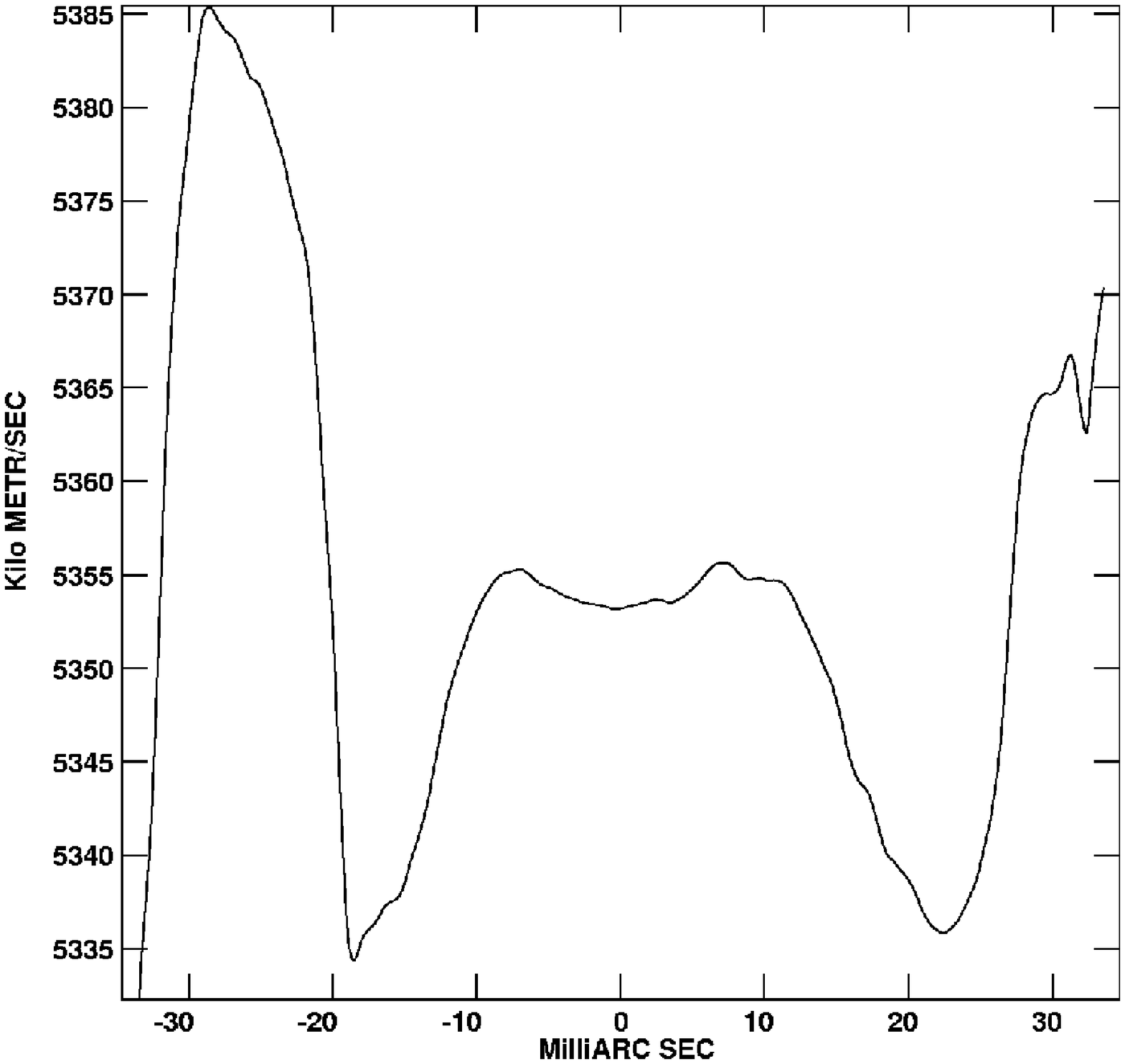}}
\end{picture}
\caption{\small The velocity structure of the northwestern region of
         Arp~220. The different velocities are taken from points along
         the yellow line of Figure 7a which lies along the major axis
         of the emission. No smoothing has been applied.}
\end{figure}

\begin{figure}
\setlength{\unitlength}{1cm}
\begin{picture}(6,8.1)
\put(-0.4,0){\includegraphics{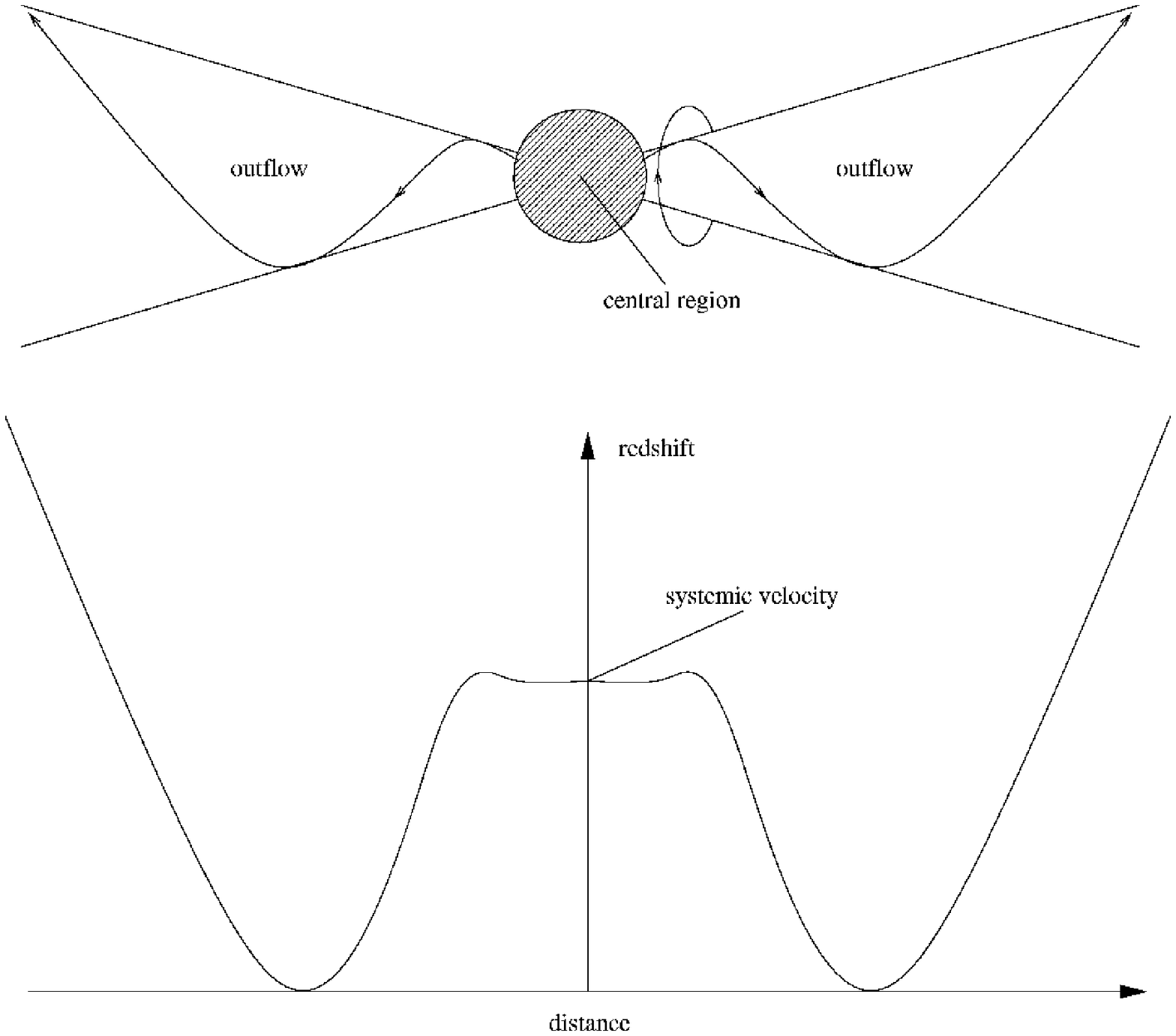}}
\end{picture}
\caption{\small The expanding helix model as an explanation for the
         VLBI northwestern velocity structure. Note that the diagram
         and the sketch of the region are not in scale.}
\end{figure}

\begin{figure}
\setlength{\unitlength}{1cm}
\begin{picture}(6,13)
\put(-0.8,0){\includegraphics{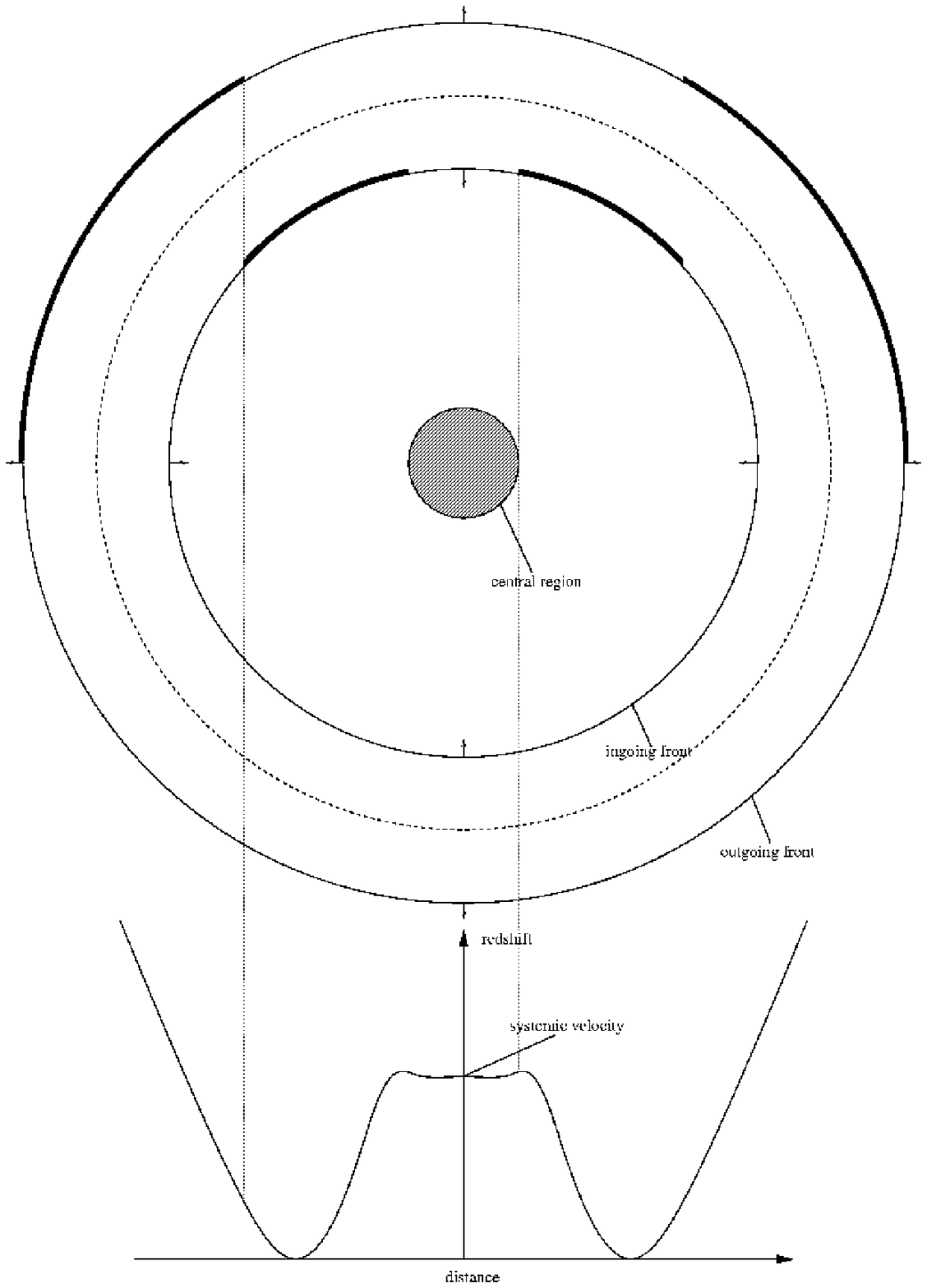}}
\end{picture}
\caption{\small The double shock front scenario as an explanation of
         the northwestern area VLBI velocity structure. Bold lined
         indicate the regions which the masers we observe come from.
         The diagram and the sketch of the region are not in scale as
         in Figure 10.}
\end{figure}

\subsection{Is There Still Room for an AGN?}

When OH megamasers were first discovered, the question of whether a
starburst or an AGN were providing the essential energy was raised.
The discovery of RSNe in Arp 220 \citep{b18} answers this question in
favour of the starburst. Luminous radio supernovae or nested
supernovae remnants have since been tentatively detected in a number
of OHMs (Mrk 273 \citep{b25}, IRAS 17028-0014 \citep{b22}, Mrk 231
\citet{b24} and IIIZw35 \citep{b12}) linking the megamaser phenomenon
with a nuclear starburst. The infrared luminosity is thought to arise
from the vast amount of dust resulting from merging phenomena heated
from young massive stars, whereas the radio continuum comes from
relativistic electrons (partially generated from young stars and
partially from supernova explosions) being accelerated by magnetic
fields, so there is a correlation between the radio and the infrared
fluxes \citep{b27}.

An interesting common feature common of OH megamasers is that the
emission comes from regions with clear velocity gradients, interpreted
as molecular discs or torii. This is evident in Arp 220 (this paper),
IIIZw35 \citep{b12}, IRAS 17028-0014 \citep{b22}, Mrk 273 \citep{b28}
and Mrk 231 \citep{b29}. The formation of such a disc (or torus) is
modeled by \citet{b30} for merging gas-rich galaxies and a similar
torus seems to be in agreement with high-resolution data of IIIZw35
\citep{b12}. Given the high column densities of these regions
(8$\times$10$^{19}$\,T$_{s}$(K)\,$<\,$N$_{H}$\,$<$\,2.4$\times$10$^{20}$\,T$_{s}$(K)\,cm$^{-2}$ for Arp 220 HI \citep{b10}) they can obscure
low luminosity AGN. Indirect evidence in the form of a radio jet in
Mrk 231 suggests that it hosts an AGN \citep{b23}. Compact hard X-ray
sources in the nuclei of Arp 220 \citep{b5} and Mrk 273 \citep{b26}
suggest that there is an active nucleus in those galaxies. The
existence of an AGN in OHM galaxies may well be a random phenomenon
not associated with their masing properties, however statistical
analysis of masing and non-masing LIRGs by \citet{b21} reveals that
LIRGs that host OH megamasers tend to be of infrared excess compared
to those that don't. This means that there is another process taking
place which contributes to the infrared flux and not the radio
continuum. This can be a hidden AGN or a very recent starburst, which
has not yet produced supernovae and relativistic electrons, OH
megamasers can work as tracer for this kind of phenomena. To
distinguish between the two possibilities one needs to model violent
and sudden starbursts to check if the time scales until it starts
contributing to the radio flux can explain the ratio of OHMs among
LIRGs. Alternatively, a large number of OHMs and host galaxies should
be observed with high sensitivity to determine if they host compact
regions, candidates for hosting an AGN.

\section{Summary}

Imaging of both the continuum and spectral line (OH) emission of
Arp~220 using the same dataset has revealed the correct relative
positions of its components. In the eastern radio nucleus the
continuum and spectral emission coincide, whereas in the western the
continuum is situated between the two spectral line emission peaks. On
VLBI scales the continuum emission is in the form of several point
sources, probably radio supernovae \citep{b18} in the same position as
the MERLIN emission.

On large (MERLIN) scales the maser emission comes from three main
regions, two western and one eastern. The eastern region has a very
clear velocity gradient, showing rotation of the nuclear disc, if we
follow the standard model \citep{b13,b10}. The velocity gradient we
measure is significantly smaller than that found using HI absorption
\citep{b10}.  This may mean that the OH rotates inside the HI on large
scales.

On smaller (VLBI) scales the OH maser emission has peculiar
characteristics. In both nuclear regions the 1667~MHz line is emitted
from regions of very small physical extent, whereas the 1665~MHz line
seems to be of a more extended nature, coming from the overall
region. Even on MERLIN scales (several hundred milliarcseconds) its
location and structure could not be reliably determined.  The most
luminous maser is situated in the northwestern region and its velocity
structure (Figure 9) implies that it either has a cone-like structure,
which could be the result of molecular outflow from the central region
or that it is produced by a pair of ingoing and outgoing shock fronts.

The issue of compact continuum structures in Arp~220 still remains
open.  Although a starburst, as exemplified by the existence of radio
supernovae, can explain Arp 220s bolometric luminosity \citep{b18},
the hard X-Ray emission spectrum and spatial distribution show that
they are unlikely to be its main power source and other alternatives
are proposed, for example several Ultra-Luminous X-Ray sources (ULX)
or one (or several) low-mass AGN \citep{b5}. The very high velocity
gradient found in the southwestern maser (Figure 7b) marks it as a
possible candidate to host a ULX or a low mass
($\sim1.7\times10^{7}\,M_{\odot}$) AGN. The sites of the radio
supernovae are of very small extent and show very faint maser
emission, so it is not feasible to study velocity gradients in these
regions. Long time monitoring and analysis of their light curves is
needed for their properties to be revealed in detail.

\section*{Acknowledgments}
MERLIN is a national facility operated by the University of Manchester
at Jodrell Bank Observatory on behalf of PPARC.

E. Rovilos would like to thank the Greek State Scholarships
Foundation, without its financial support this work would not have
been possible. We thank Al Stirling and Rob Beswick for useful
discussions.
\end{document}